\documentstyle[prl,aps,epsfig,twocolumn]{revtex}

\begin{document}
\draft

\title{Driving-Induced Symmetry Breaking in the Spin-Boson System}
\author{Holger Adam${}^1$, Manfred Winterstetter${}^1$, Milena 
 Grifoni${}^{1,2}$, and Ulrich Weiss${}^{1}$}
\address{
${}^1$ II. Institut f\"ur Theoretische Physik,
         Universit\"at Stuttgart 
         Pfaffenwaldring 57, D-70550 Stuttgart, Germany\\
${}^2$Istituto Nazionale per la Fisica della Materia,
  Corso Perrone 24, I-16152
 Genova, Italy}
\maketitle
\date{today}

\begin{abstract}
A symmetric dissipative two-state system is asymptotically
completely delocalized independent of the initial state.
We show that driving-induced localization at long times can take place
when both the bias and tunneling coupling energy are harmonically modulated.
Dynamical symmetry breaking on average
occurs when the driving frequencies are odd multiples 
of some reference frequency. This effect is {\it universal}, as it is
independent of the dissipative mechanism. Possible candidates for an
experimental observation are flux tunneling in the variable barrier rf SQUID
and magnetization tunneling in magnetic molecular clusters.
\end{abstract}


The dissipative two-state system is the simp\-lest model 
allowing the study of decoherence in a tunneling 
system\cite{Leggett1,WeissBuch}. The two states may model, e.g., 
the localized states in a tight-binding double well.
This model has found broad application to tunneling systems in 
physics and chemistry \cite{WeissBuch,Benderskii}. 
It has been used, e.g., to study the tunneling of
atoms between the tip of an atomic-force microscope and a surface 
\cite{eigler_90,louis_95}, the dyna\-mics of the magnetic flux in a 
superconducting interference device  (SQUID) \cite{han_91}, or  the 
relaxational behavior of tunneling centers in
submicron Bi-wires \cite{Golding}. Also
the low temperature dynamics of the magnetization of anisotropic 
magnetic molecular clusters can be described in terms of a two-state system 
\cite{proko}. Tunneling of magnetization has been observed
in molecular ${\rm Mn}_{12}^{}$ \cite{friedman} and 
${\rm Fe}_8^{}$ \cite{sang} nanomagnets. 

Driven dissipative systems exhibit a variety of dynamics
and opened new potentials of application \cite{MilRev}. 
 For example, the complete suppression of tunneling of a bare two-state system 
(TSS) induced by a suitably chosen monochromatic driving field \cite{Haenggi1}
 can also persist in the presence of dissipation 
\cite{Dittrich1,Dak1,Makri1}.  A variety of different 
nonequilibrium initial preparations, relevant e.g. in electron transfer,
can equivalently be described in terms of a particular time-dependent
driving force \cite{Egger4}.  An applied dc-ac field can invert
the population of donor and acceptor sites \cite{Goy1}.
Time-dependent modulation of the bias energy can be achieved, e.g.,
by an ac electric field coupled to the dipole moment of the TSS,
or, in a SQUID ring device, by a time-dependent external magnetic field 
threading the ring. 
In contrast, ``quadrupole''-like couplings induce a variation of 
barrier height and width and thus lead to a modulation of the tunneling 
matrix element (TME). 
The effects of monochromatic modulation of the TME
have been studied in Ref.~\cite{Milena5}.
Promising candidates to study the effects of 
concerted harmonic modulation of the bias {\em and} TME are 
the variable barrier rf 
SQUID ring \cite{han_91} and magnetic molecular clusters in longitudinal and 
transverse ac magnetic fields \cite{werns}.

\begin{figure}[t]
\setlength{\unitlength}{1.0cm}
\begin{center}
   \leavevmode
   \epsfig{file=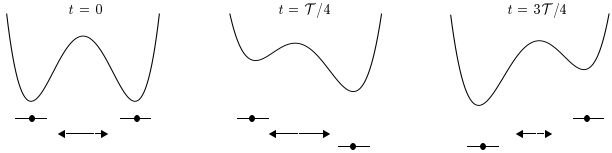,width=7.85cm,angle=0}
   \end{center}    
    \caption{\noindent The bistable potential (top) and the related TSS 
(bottom) are symbolically sketched  at different times for the case $\Omega=
      \Omega^{}_{\Delta}=\Omega^{}_{\epsilon}=2\pi/{\cal T}$.  
     At time $t=0 $, the TSS is symmetric. At time 
$t={\cal T}/4$, the right state is lower and the TME has the maximum value.
At time $t=3{\cal T}/4$, the left state is lower, but the TME
has the minimum value. Thus, relaxation towards the right state is preferred.}
    \label{fig:1}
\end{figure}

We consider the cooperative effects of two monochromatic fields modulating 
the bias energy $\hbar\epsilon$ and the TME $\hbar\Delta$.
The dissipative TSS is described by the spin-boson Hamiltonian 
\cite{Leggett1,WeissBuch,MilRev} (we set $\hbar=k^{}_{\rm B}=1$) 
\begin{equation} \label{tss}
  H (t) \,=\,-\, {\textstyle \frac{1}{2} } 
\big[ \epsilon (t) ~\sigma^{}_{z} + \Delta (t) ~\sigma_x \big]
 -{\textstyle \frac{1}{2}}\sigma_z X + H_{\rm B} \;,
\end{equation}
where  $H_{\rm B}=\sum_i[p_i^2/2m_i + m_i\omega_i^2x_i^2/2]$ represents the 
bath of bosons, and the collective variable $X=\sum_i c_i x_i$ describes the 
bath polarization.
All effects of the boson bath on the TSS are captured by the spectral density
$J( \omega ) = \frac{\pi}{2} \sum_i \frac{c^{2}_{i}}{m_{i} \omega^{}_{i}}
  \delta ( \omega - \omega^{}_{i} )$.
A dipole interaction contri\-butes to the bias energy, whereas a quadrupole 
modulation  results in an exponential modification of the TME,
\begin{equation}
\Delta (t) = \Delta^{}_{0} \; \exp[A_\Delta
\sin(\Omega^{}_{\Delta} t)] \; , \ \ \ 
\epsilon (t) = \epsilon^{}_{1} \sin(\Omega^{}_{\epsilon} t) \;.
\label{definition}
\end{equation}
In the absence of driving, $A_\Delta=0,\; \epsilon_1^{}=0$, the equilibrium 
state reached at long times has the localized states $|R\rangle$ and 
$|L\rangle$ occupied with equal probability. 
If either the bias or the coupling energy is modulated, the left-right
symmetry is dynamically broken. However, on average over a period,
one finds asymptotically again equal occupation of both states.
When both parameters are modulated with commensurable frequencies 
($n,\;m$ integer)
\begin{equation}
\Omega^{}_{\epsilon} = m \Omega \ \ \mbox{and}
\ \ \Omega^{}_{\Delta} = n \Omega\;,
\label{factor}
\end{equation} 
the Hamiltonian (\ref{tss}) is periodic
(we assume ${\cal T} = 2\pi/\Omega$ to be the smallest period in common). 
We find that equal occupation on average is reached at long times when $n$ or 
$m$ is even. However, when both $n$ and $m$ are odd, the right-left  symmetry 
is broken even on average. The symmetry breaking is maximal when $n=m=1$.
The phenomenon that the occupation of one of the two states is preferred on
average at long times is schematically sketched and explained in  
Fig.~\ref{fig:1}. In the following, we give a quantitative study of this 
effect.

Suppose that the TSS has been prepared at time zero
in the state $|R\rangle$ $(\sigma^{}_{z}=+1)$ with the
bath in thermal equilibrium.
The dynamical quantity of interest is then the population difference
$\langle \sigma^{}_{z} \rangle^{}_{t}\equiv P(t) =P_R(t) -P_L(t)$ 
for this factorizing initial condition.
Since the bath is harmonic, it can be traced out exactly.
Then,  $P(t)$ is expressed as a double path integral over the forward and 
backward spin paths $\sigma(\tau)$ and $\sigma^{\prime}_{}(\tau)$ which are
piecewise 
\noindent
constant with values $\pm 1$.
The effects of the environment are in an influence
functional which introduces nonlocal correlations 
between different path segments \cite{WeissBuch}.
It is convenient to switch to the path combinations
$\eta(\tau)=[\sigma(\tau)+\sigma^{\prime}(\tau)]/2$ and 
$\xi(\tau)=[\sigma(\tau)-\sigma^{\prime}(\tau)]/2$.
Then, $P(t)$ is expressed in terms of the double path sum
\[
P (t) = \int {\cal D}\xi \, {\cal D}\eta \,{\cal A}[\xi,\eta]
\exp\Big[\Phi{'}_{FV}[\xi]+ 
i\, \Phi{''}_{FV} [\xi,\eta]\Big]  \; , 
\]
where ${\cal A}$ is the path weight in the absence of the bath 
coupling, and the influence function is
\begin{eqnarray}
\Phi^{}_{FV} [\xi,\eta] &=&  \!\int^{t}_{0}\! d t^{}_{2}\! 
\int^{t^{}_{2}}_{0} \! d t^{}_{1} \, 
{\dot \xi} (t^{}_{2}) \nonumber\\
&& \!\! \times \big[ Q^{\prime}_{} (t^{}_{2} - t^{}_{1}) {\dot \xi} (t^{}_{1})
+ i Q^{\prime \prime}_{} (t^{}_{2} - t^{}_{1}) {\dot \eta}
(t^{}_{1}) \big] \; .
\label{influence}
\end{eqnarray}  

For the Ohmic spectral density 
$J(\omega)= 2 \pi \alpha \,\omega \,e^{-\omega/\omega_{c}}_{}$,
where $\alpha$ is the dimensionless Ohmic coupling, the kernel takes for 
times $\omega_c\tau \gg 1$ the so-called {\em scaling} form  
\begin{equation}
Q(\tau) = 2 \alpha \ln{[ ( \beta \omega^{}_{c}/ \pi)
\sinh{(\pi \tau/ \beta)} ]} \,+\,i\, \pi \alpha \, {\rm sgn} \tau \;.
\label{scaling}
\end{equation}

It is convenient to write $P(t)= P^{}_{s} (t) + P^{}_{a}(t)$, 
where $P^{}_{s} (t)$ and $P^{}_{a} (t)$ are the
symmetric and antisymmetric parts under inversion of the bias 
$\epsilon \rightarrow - \epsilon$. 
Since $\lim^{}_{t \rightarrow \infty} P^{}_{s} (t) = 0$, the 
asymptotic value is determined by the 
antisymmetric part $P^{}_{a} (t)$.
In the {\it absence} of the biasing force, $\epsilon_1^{}= 0$,
we have $P^{}_{a} (t)=0$ for all $t$. For {\em any}  
initial preparation, we then have 
 $\lim^{}_{t \rightarrow \infty} P(t) =0$,
 i.e., the TSS is completely delocalized at long times. 
Similarly, if we choose $A_\Delta=0$ and
$\epsilon^{}_{1} \not= 0$, the time average  
$\langle \sin{\zeta(t^{}_{2},t^{}_{1})} \rangle^{}_{\cal T}$ over
a period $2\pi/\Omega_\epsilon$ of the external field vanishes. Thus, 
on {\it average}, the TSS is again delocalized,
 i.e., $\overline{P}^{}_{\infty} \equiv \lim^{}_{t \rightarrow \infty} 
\langle P(t) \rangle^{}_{\cal T}=0$.   
However, the constructive interference between diagonal and off-diagonal 
driving, i.e. $A_\Delta \not=0$ and $\epsilon_1\not=0$,
can lead to driving-induced symmetry breaking, as qualitatively illustrated 
in Fig. \ref{fig:1}. We find $\overline{P}^{}_{\infty}\not=0$ 
for $\Omega^{}_{\epsilon}=\Omega^{}_{\Delta}$, and the tight-binding site 
preferred is determined by the relative sign of $A_\Delta$ and $\epsilon_1^{}$.

The evolution of $P(t)$ can not 
be given in analytic form apart from special limits.
Numerical methods cover short to intermediate times, but they become 
costly or even inadequate if one is interested in the asymptotic behavior. 
\begin{figure}[htb]
 \begin{center}
   \leavevmode
   \epsfig{file=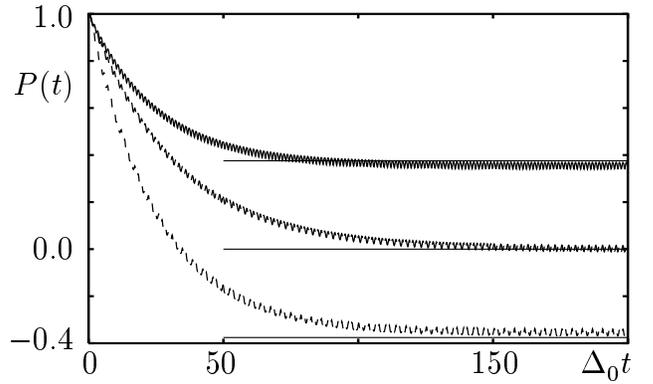,height=5.1cm,angle=0}
\end{center}
\caption{\label{fig:2}
Plots of the population difference $P(t)$ for $\alpha=1/2$ as calculated from 
the expression (\ref{exact}). The driving-induced symmetry breaking
phenomenon is clearly visible. The parameters are
$T=0.05 \Delta^{}_{0}$, $\omega^{}_{c}=50 \Delta^{}_{0}$, 
    $\epsilon^{}_{1} = 5 \Delta^{}_{0}$, and 
    $\Omega^{}_{\epsilon}=\Omega^{}_{\Delta}=5 \Delta^{}_{0}$.
The upper curve describes the case $A_\Delta =   0.5 $,
and the lower curve the case $A_\Delta=-0.5$.
The horizontal lines follow from Eq.~(\ref{asymptotic2}) with 
(\ref{asymptotic}). 
The middle curve corresponds to a time-independent TME $(A_\Delta=0)$. In this
case, the right/left symmetry is asymptotically restored.}
\end{figure}

For the special case $\alpha=1/2$ of the Ohmic coupling, however, 
the dynamics can be found in analytic form in the scaling limit
(\ref{scaling}).
Using the concept of collapsed blips \cite{WeissBuch}, we find
\begin{eqnarray}  \nonumber
P^{}_{s}(t) &=& e^{- G(t,0)} \; ,
\qquad
G (t^{}_{2},t^{}_{1}) = \frac{\pi}{2\omega_c}
\int^{t^{}_{2}}_{t^{}_{1}} \!ds\,  \Delta^{2}_{} (s) \;, \\[1mm]  \label{exact}
P^{}_{a} (t) &=& \int^{t}_{0} \! d t^{}_{2} 
\int^{t^{}_{2}}_{0} \!d t^{}_{1} \, 
e^{- G(t,t^{}_{2})}_{} 
\Delta ({t^{}_{2}}) \,  \Delta ({t^{}_{1}}) \; \\  \nonumber
&& \times\; e^{- Q^{\prime}_{} (t^{}_{2} - t^{}_{1})} 
\sin{\big[ \zeta (t^{}_{2},t^{}_{1}) \big] } \; 
e^{-G (t^{}_{2},t^{}_{1})/2}_{} \; , 
\end{eqnarray}
with
$\zeta (t^{}_{2},t^{}_{1}) = \epsilon^{}_{1} \int^{t^{}_{2}}_{t^{}_{1}}\!ds\, 
\sin(\Omega^{}_{\epsilon} s) \, .$
The relaxation of $P(t)$ is shown for different $A_\Delta^{}$ in
Fig.~2. The superimposed  oscillation is due to the oscillatory bias.
The middle curve is for $A_\Delta^{}=0$. In this case, 
the TSS is asymptotically delocalized on average.
Symmetry breaking only occurs when both
the bias and the intersite coupling are monochromatically modulated
(upper and lower curve).
The results depicted in Fig.~2 correspond to
the case $\Omega^{}_{\Delta}=\Omega^{}_{\epsilon}$.
Evidently, the sign of $P^{}_{a} (t)$ depends on the sign of $A_\Delta$.

In the general case, the dynamics of the driven system can be described by the 
exact master equation \cite{Gri96}
\begin{equation}
{\dot P} (t) = \int^{t}_{0} d t^{\prime}_{} 
\; \big[ K^{}_{a} (t,t^{\prime}_{}) -
K^{}_{s} (t,t^{\prime}_{}) ~P(t^{\prime}_{}) \big] \; .
\label{niba}
\end{equation}
The familiar {\em noninteracting-blip
approxi\-mation} (NIBA) describes the dynamics correctly for high 
temperature and/or strong damping \cite{Leggett1,WeissBuch}.
The NIBA kernels read
\begin{equation}
K^{}_{s/a} (t, t^{\prime}{}) = 
\,e^{-Q'(t-t')}_{} B_{s/a}^{} [ Q^{\prime \prime} (t-t')]
L^{}_{s/a} (t, t^{\prime}{}) \; , \label{kernel}
\end{equation}
where $B_s(z) = \cos(z)$ and $B_a(z)=\sin(z)$.
The effects of the driving forces are captured by
\begin{equation}
L^{}_{s/a} (t,t^{\prime}_{}) =  
\Delta (t) \Delta (t^{\prime}_{})
B_{s/a}{\big[ \zeta (t,t^{\prime}_{}) \big]}\; .
\end{equation}
The equilibrium state for a TSS with static bias $\epsilon$ is 
$P^{}_{\infty} = [k_{+}^{}(\epsilon) -
k_{-}^{}(\epsilon)]/[k_{+}^{}(\epsilon) +
k_{-}^{}(\epsilon)]$, where $k_{\pm}^{}(\epsilon)$ is the forward/backward
rate for the static case. One has 
$k_{\pm}^{}(\epsilon) = (\Delta^{2}_{0}/4)\int^{\infty}_{-\infty}
dt \,\exp[\pm i\epsilon t - Q(t)]$, obeying detailed balance, 
$k_{-}^{}(\epsilon) = \exp(- \beta \epsilon) k_{+}^{}(\epsilon)$.
In the scaling limit, the Ohmic 
forward rate reads \cite{WeissBuch}
\begin{equation}
k_{+}^{}(\epsilon)\! = \! \frac{1}{4}
\frac{\Delta^{2}_{0}}{\omega^{}_{c}}  
\Big( \frac{\beta \omega^{}_{c}}{2 \pi} \Big)^{\!1\!-\!2\alpha}
\frac{|\Gamma(\alpha \!+\! i \beta \epsilon/2 \pi)|^{2}}
{\Gamma(2\alpha)}\; e^{\beta \epsilon/2}_{} \; .
\label{rate}
\end{equation}
For the driven TSS, Eqs.~(\ref{tss}) -- (\ref{factor}), the average rate is 
\[
\overline{k}_{\pm}\,=\, \frac{\Omega}{2\pi}\! 
\int_{0}^{2\pi/\Omega} \!\!\!dt \!\int_{0}^{\infty} \!\!d\tau \; \frac{1}{2}
\big[\,K^{}_{s} (t,t\!-\!\tau)\!\pm\! K^{}_{a} (t,t\!-\!\tau)\,\big] \;.
\]
Next, we expand the bias phase terms 
$e^{\mp i(\epsilon_1^{}/
\Omega_{\epsilon})\cos(\Omega_{\epsilon} t)}_{}$ and
$e^{\pm i(\epsilon_1/\Omega_\epsilon)\cos[\Omega_\epsilon(t-\tau)]}_{}$
in series in $J$-Bessel functions, and the oscillatory factors of the tunneling
coupling $e^{A_\Delta\sin(\Omega_\Delta t)}_{}$ and
$e^{A_\Delta \sin[\Omega_\Delta(t-\tau)]}_{}$ into series in $I$-Bessel
functions. We then find the average rate $\overline{k}_\pm^{}$ in the form
\begin{eqnarray}
\overline{k}_\pm^{} &=&  \! \sum^{\infty}_{\ell,r,s=-\infty} \!\!   
e^{i\pi(n/m-1)(r+s)/2}_{} \; 
k_{\pm}^{} [\,(m\ell \!-\!ns)\Omega\,]  \nonumber\\
&& \!\!\! \times\; I^{}_{r} (A_\Delta) I^{}_{s} (A_\Delta) J^{}_{\ell} 
\Big(\frac{\epsilon^{}_{1}}{m \Omega}\Big)
J^{}_{\ell-(r+s)n/m } \Big(\frac{\epsilon^{}_{1}}{m\Omega}\Big) \; .
\label{asymptotic}  
\end{eqnarray} 
The individual channels in (\ref{asymptotic}) describe tunneling under 
emission and absorption of a fixed number of quanta with frequencies which are
multiples of the frequency $\Omega$.
The channel weights are given in terms of $I$ and $J$ Bessel functions.
The exclusive channel rates are expressed in terms of the rates $k_\pm^{}$
for a static bias $\epsilon = (m\ell-ns)\Omega$.

The mean asymptotic position is then found as
\begin{equation}
\overline{P}^{}_{\infty}= (\overline{k}_{+}^{}- \overline{k}_{-}^{})
/(\overline{k}_{+}^{} + \overline{k}_{-}^{}) \; . \label{asymptotic2}
\end{equation}
Substituting Eq.~(\ref{asymptotic}) and the symmetry 
relation $k_{-}(\epsilon) = k_{+}(-\epsilon)$, and using the properties 
$J_{-n}(z) = (-1)^n J_n(z)$ and $I_{-n}(z) = I_n(z)$,  we find  
$\overline{P}_\infty =0$ when $n$ or $m$ is even, whereas 
$\overline{P}_\infty \neq 0$ when both $n$ and $m$ are odd.

The predictions following from (\ref{asymptotic2}) are shown  in 
Fig.~\ref{fig:2} (horizontal lines) in comparison with the
exact $P(t)$ (oscillating curves). We have chosen $\alpha=1/2$
to allow for a direct comparison with the
predictions of the exact solution (\ref{exact}). Evidently, 
the asymptotic position value is very accurately
reproduced by the NIBA.

\begin{figure}[b!]
\vspace*{-2mm}
\begin{center}
\leavevmode
\epsfig{file=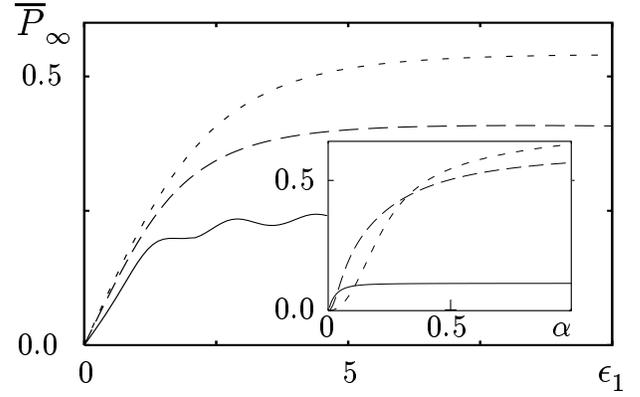,height=5.2cm,angle=0}
\end{center}  
\vspace*{-0.1cm}    
\caption{\label{fig:3}  
The mean asymptotic value $\overline{P}{}_{\infty}$
is shown as a function of the bias amplitude $\epsilon^{}_{1}$  for 
Ohmic couplings $\alpha=0.1$ (full curve), $\alpha=1/4$ (long-dashed curve), 
and $\alpha=1/2$ (short-dashed curve).
The parameters are $A_\Delta = 0.5 $, 
$\Omega^{}_{\epsilon}=\Omega^{}_{\Delta}=\Delta^{}_{0}/2$,
$T=\Delta^{}_{0}$, and $\omega^{}_{c}=50 \Delta^{}_{0}$.
The inset shows $P^{}_{\infty}$ versus $\alpha$ for temperatures 
$T=0.1\Delta^{}_{0}$ (short-dashed curve), $T=\Delta^{}_{0}$ (dashed curve),
and $T=10 \Delta^{}_{0}$ (full curve).
The bias amplitude is $\epsilon^{}_{1}=10\Delta^{}_{0}$.}
\end{figure}

In Fig. \ref{fig:3} we show $\overline{P}^{}_{\infty}$ as a function of the
bias amplitude $\epsilon^{}_{1}$ for different coupling strengths.
The full curve ($\alpha=0.1$) exhibits oscillations. The oscillations
are already damped out in the curves for the higher damping values displayed. 
Interestingly, the symmetry breaking effect grows with increasing
$\alpha$ and increasing $\epsilon^{}_{1}$. 
For large bias amplitude, in the investigated bias regime,
  $\overline{P}^{}_{\infty}$  saturates. The inset
depicts $\overline{P}^{}_{\infty}$ versus $\alpha$ for 
$\epsilon^{}_{1}=10\Delta^{}_{0}$ and different temperatures. 
For weak coupling $\alpha\ll 1$, 
\linebreak
\begin{figure}[htb]
\vspace{-4mm}  
\begin{center}
   \leavevmode
   \epsfig{file=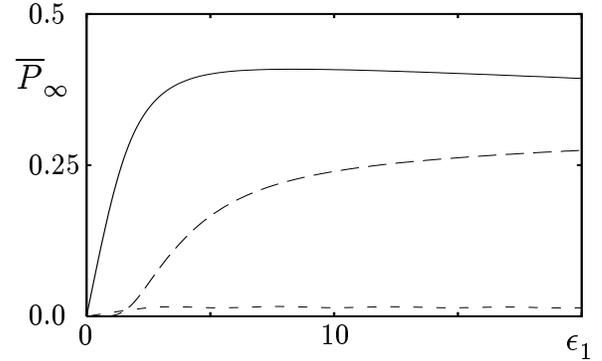,height=4.8cm,angle=0}
\end{center}
\caption{\noindent Plots of $\overline{P}^{}_{\infty}$ versus bias 
$\epsilon^{}_{1}$ for 
$\Omega^{}_{\Delta}=\Omega^{}_{\epsilon}=\Delta^{}_{0}/2$ (full curve),
$\Omega^{}_{\Delta}=3\Omega^{}_{\epsilon}=\Delta^{}_{0}/2$ 
(long-dashed curve), and
$\Omega^{}_{\epsilon}=3\Omega^{}_{\Delta}=\Delta^{}_{0}/2$ 
(short-dashed curve). The other parameters are     
$\alpha=1/4$, $A_\Delta =0.5$,
$T=\Delta^{}_{0}$ and $\omega^{}_{c}=50 \Delta^{}_{0}$.}
 \label{fig:4} 
\end{figure}
\noindent
the symmetry breaking becomes more effective with increasing
temperature. For stronger coupling, on the contrary, it is more pronounced
the lower the temperature.

In Fig.~\ref{fig:4}, we show $\overline{P}^{}_{\infty}$ for different ratios
$\Omega_\epsilon/\Omega_\Delta$. The full and long-dashed curves represent
the cases $\Omega^{}_{\epsilon}=\Omega^{}_{\Delta}$ and 
$\Omega_\Delta = 3\Omega_\epsilon$. 
In both cases, the symmetry breaking is clearly evident.
When the bias frequency is three times the barrier modulation
frequency, the symmetry breaking effect is strongly reduced
(short-dashed curve). The different behavior is due to the oscillatory and
smooth behavior of the $J_n(z)$ and $I_n(z)$ functions, respectively.

We believe that the phenomenon of dynamical symmetry breaking 
can be seen, e.g, in Josephson systems.
A promising candidate is a superconducting quantum interference device
in which the tunneling coupling and bias energy can be varied independently
\cite{han_91}.
In the variable barrier rf SQUID, the single Josephson junction in a standard
rf SQUID ring with inductance $L$ is replaced by a superconducting loop with 
inductance $L_{dc}$ interrupted by two junctions. 
In the limit $L_{dc}/L\to 0$,
the  2D Josephson potential $V(\Phi,\Phi_{dc})$ goes over into the 1D 
potential of a standard rf SQUID with adjustable critical 
current $I_c(\Phi_{x\,dc})$. Within the RSJ model, 
the deterministic equation of motion of the rf SQUID for the flux variable
$\phi= 2\pi\Phi/\Phi_0=2e\Phi$ is
$C\ddot{\phi} + \dot{\phi}/R + 4e^2_{}(\partial/\partial\phi)U(\phi)  =0$.

The rf~SQUID potential is given by ($U_0 = 1/4e^2_{}L$)
\vspace{-1mm}  
\begin{equation}\label{pot}
U(\phi) = U_0[\,{\textstyle\frac{1}{2}} (\phi-\phi_x\,)^2 - 
\beta_L(\phi_{x\,dc}) \cos(\phi)\,],  \\[-1mm]
\end{equation}
where $\beta_L(\phi_{x\,dc}) = 2e L I_c \cos(\frac{1}{2} \phi_{x\,dc})$.
For an applied flux $\phi_x=\pi$ and $\beta_L>1$, the potential is symmetric 
about $\phi=\pi$ with two degenerate minima at $\phi=\pi \pm \phi_m$,
where $\phi_m$ is the smallest positive root of
$\phi_m -\beta_L \sin \phi_m =0$. The barrier height is 
$U_b = U_0[\beta_L - \frac{1}{2}\phi_m^2 -\beta_L\cos\phi_m]$, and the squared
well frequency is $\omega_0^2 = (1 - \beta_L\cos\phi_m)/LC$. 

With the modulation $\phi_x = \pi + \phi_1^{} \sin(\Omega_\epsilon t)$,
where $\phi_1^{}\ll \phi_m$, the system is effectively described at 
low $T$ by Eq.~(\ref{tss}) with the bias variation (\ref{definition}), and
amplitude $\epsilon_1^{} = \phi_m\phi_1^{}/2e^2 L$. 
The parameter $v \equiv U_b/\omega_0$ can be 
varied {\em independently} of the bias by varying the flux $\phi_{x\,dc}$
applied to the small loop.
For $\beta_L \gg 1$,  the tunneling length $2\phi_m$, and hence the
damping parameter $\alpha=(\phi_m/\pi)^2 R_Q/R$ 
($R_Q = \pi/2e^2= 6.5\;k\Omega$), is approximately constant under variation of
$\phi_{x\,dc}$. Harmonic modulation of the WKB exponent $v$ of the TME
$\Delta \propto \exp(-c v)$, where $c$ depends on the barrier shape,
can now be achieved by driving $\phi_{x\,dc}$. Thus one can 
experimentally reach the situation described by Eq.~(\ref{definition}). 
Typical realizable junction and loop parameters are 
$L \approx 200\,$pH, $C\approx 100\,$fF, $\beta_L\approx 5$, 
and $R/R_Q$ ranging from 0.5 to 10. 
This results in values for $\alpha$ ranging from 
$0.1$ to $2$, barrier height $U_b \approx 1 \, \ldots 10 \,$K,
well frequency $\omega_0^{}\approx 10^{10} \ldots10^{11}\,$s$^{-1}$, and the
tunneling matrix element in the MHz regime.
We expect that the symmetry breaking effect can be observed at temperatures 
about $10\,$ mK and driving frequencies $\Omega_\Delta$ and 
$\Omega_\epsilon$ in the $1\,\ldots\,10\,$MHz regime.

A giant spin or nanomagnet is another possible candidate for the 
observation of the above symmetry breaking phenomenon.
In these systems, the
bias can be tuned by a magnetic field along the easy axis. It has been 
demonstrated that the TME of the ${\rm Fe}_8^{}$ molecular system can be 
tuned by a transverse magnetic field \cite{werns},
as theoretically predicted \cite{garg}. Therefore, it seems feasible to
operate nanomagnets in the appropriate parameter regime.

In summary, we have found that dynamical symmetry breaking 
can be induced by a concerted modulation of the bias
and tunneling energy of a symmetric TSS. 
Because of constructive interference, dwelling in one of the  localized states
is preferred on average. This effect is robust since it is
largely independent of the dissipative mechanism. 
Also, the dependence on coupling strength and 
temperature is interesting. For weak coupling, the asymptotic
mean position increases with temperature, whereas for strong
coupling it shows opposite behavior. 

Readily, the results can be extended to quantum transport in 
a tight-binding (TB) lattice. In the incoherent tunneling regime, the  
asymptotic current is proportional to the  asymptotic mean position of the 
TB system \cite{MilRev}. Upon imposing concerted driving of
bias and TME, it is possible to extract 
a finite current out of an (on average) undirectional force.
Thus, the system constitutes a novel realization of a discrete quantum 
Brownian rectifier \cite{GoyTB}.

We wish to thank F. Chiarello for valuable discussions on 
the variable barrier rf SQUID, and an anonymous referee for pointing out
the possible relevance to magnetic molecular clusters. We acknowledge  
support by the Deutsche Forschungsgemeinschaft (DFG)
through the Sonderforschungsbereich 382,  by the Max Planck
Institut f\"ur Festk\"orperforschung in Stuttgart, and by the Istituto 
 Nazionale per la Fisica della Materia (INFM) under the PRA-QTDM
 Programme.


\begin{references}

\bibitem{Leggett1}A.J. Leggett {\em et al.}, 
Rev. Mod. Phys. {\bf 59},~1 (1987).
\bibitem{WeissBuch} U. Weiss, {\em Quantum Dissipative Systems}
    (World Scientific, Singapore, Second Edition, 1999).
\bibitem{Benderskii} A. Benderskii {\em et al.},
  Adv. Chem. Phys. {\bf 88}, 1 (1994).
\bibitem{eigler_90} D.M. Eigler and E.K. Schweizer, Nature {\bf 344}, 524
  (1990).
\bibitem{louis_95} A.A. Louis and J.P. Sethna, Phys. Rev. Lett. {\bf 74},
  1363 (1995).
\bibitem{han_91} S. Han, J. Lapointe, and J.E. Lukens,
  Phys. Rev. Lett. {\bf 66}, 810 (1991); Phys. Rev. B {\bf 46}, 6338 (1992).
\bibitem{Golding} B. Golding {\em et al.}, 
Phys. Rev. Lett. {\bf 68}, 998 (1992).
\bibitem{proko} N.V. Prokof'ev and P.C.E. Stamp, Phys. Rev. Lett. {\bf 80}, 
5794 (1998).
\bibitem{friedman} J.R. Friedman {\em et al.}, Phys. Rev. Lett. {\bf 76},
3830 (1996).
\bibitem{sang} C. Sangregorio {\em et al.}, Phys. Rev. Lett. {\bf 78},
4645 (1997).
\bibitem{MilRev} M. Grifoni and P. H\"anggi, Phys. Rep. {\bf 304}, 229 
(1998).
\bibitem{Haenggi1} F. Grossmann {\em et al.},
Phys. Rev. Lett. {\bf 67}, 516 (1992);
F. Grossmann and P. H\"anggi,
 Europhys. Lett. {\bf 18}, 571 (1992).
\bibitem{Dittrich1} T. Dittrich {\em et al.},
   Europhys. Lett. {\bf 22}, 5 (1993).    
\bibitem{Dak1} Yu. Dakhnovskii, Phys. Rev. B {\bf 49}, 4649 (1994); 
         Yu. Dakhnovskii, Ann. Phys. {\bf 230}, 145 (1994).
\bibitem{Makri1} D.E. Makarov and N. Makri,
    Phys. Rev. E {\bf 52}, 5863 (1995).
\bibitem{Egger4} A. Lucke {\em et al.}, J. Chem. Phys. {\bf 107}, 8397 (1997).
\bibitem{Goy1} I.A. Goychuk {\em et al.}, 
Chem. Phys. Lett. {\bf 253}, 428 (1996).
\bibitem{Milena5} M. Grifoni, Phys. Rev. E {\bf 54}, R3086 (1996).
\bibitem{werns} W. Wernsdorfer and R. Sessoli, Science {\bf 284}, 133 (1999).
\bibitem{Gri96} M. Grifoni, M. Sassetti, and U. Weiss, Phys. Rev. E {\bf 53},
R2033 (1996).
\bibitem{GoyTB} S. Yukawa, M. Kikuchi, G. Tatara and M. Matsukawa,
 J. Phys. Soc. Japan {\bf 66}, 2953 (1997); 
I. Goychuk, M. Grifoni and P. H\"anggi, Phys. Rev. Lett.
 {\bf 81}, 649 (1998); I. Goychuk and P. H\"anggi, Europhys. Lett. 
 {\bf 43}, 503 (1998).

\bibitem{garg} A. Garg, Europhys. Lett. {\bf 22}, 205 (1993).

\end{references}
\end{document}